\documentclass[journal=nalefd,manuscript=letter]{achemso}

\usepackage{makeidx}\makeindex
\usepackage{amsmath}
\usepackage{amssymb}
\usepackage{amsthm}
\usepackage{graphicx} 
\usepackage{color}
\usepackage{bm}

\usepackage{achemso}
\setkeys{acs}{articletitle = true}

\title{Direct Observation of the Zigzag Edge States of a Supramolecular Diatomic Kagome Lattice}

\author{Ryohei Nemoto}
\affiliation{Research Center for Materials Nanoarchitectonics (MANA), National Institute for Materials Science, 1-1, Namiki, Tsukuba, Ibaraki 305-0044, Japan}
\alsoaffiliation{Department of Physics, School of Science, Institute of Science Tokyo, 2-12-1, Ookayama, Meguro-ku, Tokyo 152-8551, Japan}

\author{Xiangzhi Meng}
\affiliation{Institut f\"{u}r Experimentelle und Angewandte Physik, Christian-Albrechts-Universität, 24098 Kiel, Germany}

\author{Behzad Mortezapour}
\affiliation{Institut f\"{u}r Experimentelle und Angewandte Physik, Christian-Albrechts-Universität, 24098 Kiel, Germany}

\author{Alexander Weismann}
\email{weismann@physik.uni-kiel.de}
\affiliation{Institut f\"{u}r Experimentelle und Angewandte Physik, Christian-Albrechts-Universität, 24098 Kiel, Germany}

\author{Saya Nakano}
\affiliation{Department of Physics, Nara Women's University, Kitauoyanishi-machi, Nara 630-8506, Japan}

\author{Masahisa Tsuchiizu}
\affiliation{Department of Physics, Nara Women's University, Kitauoyanishi-machi, Nara 630-8506, Japan}

\author{Ryuichi Arafune}
\affiliation{Research Center for Materials Nanoarchitectonics (MANA), National Institute for Materials Science, 1-1, Namiki, Tsukuba, Ibaraki 305-0044, Japan}

\author{Noriaki Takagi}
\affiliation{Graduate School of Human and Environmental Studies, Kyoto University, Kyoto 606-8501, Japan}

\author{Sho Nakamura}
\affiliation{School of Science and Technology, Kwansei Gakuin University, Sanda, Hyogo 669-1330, Japan}

\author{Katsunori Wakabayashi}
\affiliation{Research Center for Materials Nanoarchitectonics (MANA), National Institute for Materials Science, 1-1, Namiki, Tsukuba, Ibaraki 305-0044, Japan}
\alsoaffiliation{School of Science and Technology, Kwansei Gakuin University, Sanda, Hyogo 669-1330, Japan}

\author{Rie Suizu}
\email{rsuizu@cc.saga-u.ac.jp}
\affiliation{Synchrotron Light Application Center, Saga University, 1 Honjo, Saga, 840-8502, Japan}
\alsoaffiliation{Department of Chemistry and IRCCS, Nagoya University, Furo-cho, Chikusa-ku, Nagoya 464-8602, Japan}
\alsoaffiliation{Japan Science and Technology Agency (JST), PRESTO, 4-1-8 Honcho, Kawaguchi, Saitama, 332-0012, Japan}

\author{Richard Berndt}
\affiliation{Institut f\"{u}r Experimentelle und Angewandte Physik, Christian-Albrechts-Universität, 24098 Kiel, Germany}

\author{Takashi Uchihashi}
\email{UCHIHASHI.Takashi@nims.go.jp}
\affiliation{Research Center for Materials Nanoarchitectonics (MANA), National Institute for Materials Science, 1-1, Namiki, Tsukuba, Ibaraki 305-0044, Japan}
\alsoaffiliation{Graduate School of Science, Hokkaido University, Kita-10 Nishi-8, Kita-ku, Sapporo 060-0810, Japan}

\author{Kunio Awaga}
\affiliation{National Institute of Technology (KOSEN), Toyota College, 2-1 Eiseicho, Toyota, Aichi 471-8525, Japan}
\alsoaffiliation{Department of Chemistry and IRCCS, Nagoya University, Furo-cho, Chikusa-ku, Nagoya 464-8602, Japan}
\begin{document}



\section*{Abstract}
\textcolor{black}{
Lattice geometry plays a fundamental role in the behavior of Bloch electrons in a crystal. 
The diatomic Kagome lattice, an extension of the honeycomb and Kagome lattices, is predicted to give rise to emergent and topological phenomena, but its experimental investigation has been limited thus far. 
Here, we fabricate a diatomic Kagome lattice through self-assembly of a triptycene derivative with phenazine moieties (Trip-Phz)---a $\mathrm{C_3}$-symmetric, non-planar $\pi$-conjugated molecule.
Our scanning tunneling microscopy (STM) observations show that Trip-Phz forms a highly ordered diatomic Kagome lattice terminated by zigzag-type edges on the Pb(111) surface. 
Combined STM measurements and \textcolor{black}{tight-binding} calculations provide direct evidence for the existence of the edge states that correspond to those of graphene.
These states are topological edge states dictated by the quantization of the Zak phase and the bulk-edge correspondence. 
This work reveals an ideal platform for exploring quantum materials with unique lattice geometries using supramolecular technology.
}

\section*{Keywords}
Diatomic-Kagome lattice, Edge states, Zak phase, Self assembly, Non-planar $\pi$-conjugated molecules, Scanning tunneling microscopy

\newpage

Lattice geometry profoundly affects the behavior of itinerant electrons in a crystal and is central to the modern research on quantum materials and technologies \cite{Leykam_2018_3118, Springer_2020_3405,Smirnova_2020_3758}. 
The honeycomb and Kagome lattices, the latter of which consists of corner-sharing triangles with enclosed hexagons, are two-dimensional (2D) lattices that have been studied the most intensively in the past decade \cite{Geim_2009_3690, Polini_2013_1695,Yin_2022_3757}. 
Tight-binding (TB) calculations predict the formation of linearly intersecting bands (Dirac bands) for both lattices and a nearly dispersionless band (flat band) originating from geometric frustration for the Kagome lattice \cite{Ando_2005_3685,Guo_2009_2959}. 
These remarkable band structures possess nontrivial topology and \textcolor{black}{can be subject to strong electron correlation effects due to a vanishing kinetic energy}, leading to rich emergent phenomena such as quantum anomalous Hall effect and superconductivity \cite{Haldane_1988_3007,Kane_2005_1682,Novoselov_2005_184,Tang_2011_2934,Ye_2018_2941,Li_2018_3733,Ortiz_2020_3751,Jiang_2021_2938}.

The honeycomb lattice can be extended in such a way that triangles with an internal coupling constant $t_0$ are placed at the lattice nodes while they are coupled to their nearest neighbors with $t_1$ (see Fig.~1(a)). 
This can also be viewed as an extended Kagome lattice, where the vertices of the triangular units are connected by \textcolor{black}{a diatomic pair  (indicated by the dashed orange line in Fig.~1(a))}.
This diatomic Kagome lattice, also called the star lattice because of its shape, can be regarded as a hybrid of the two basic lattices \cite{Zheng_2007_3433,Mizoguchi_2019_3152,Jiang_2021_3354}. 
TB calculations predict the formation of four Dirac bands and two flat bands, the topology of which is switched by tuning the ratio $t_0/t_1$ \cite{Shuku_2018_2707,Hu_2022_3748,Hu_2023_3247,Shuku_2023_3286}. 
Recently, the diatomic Kagome lattice has attracted theoretical attention and is predicted to exhibit novel quantum states such as second-order topological states and excitonic insulators \cite{Guo_2020_3749,Sethi_2021_2943,Hu_2022_3748,Sethi_2023_3747,Hu_2023_3247}. 
However, its experimental investigation has been limited so far \cite{Telychko_2021_3094,Yin_2024_3409,Hu_2023_3247}.
Particularly, the one-dimensional edge states corresponding to those of graphene \cite{Fujita_1996_3679, Nakada_1996_3680,Ryu_2002_3687,Kobayashi_2005_3678,Li_2021_3710}, which connect the projected Dirac points in momentum space and exhibit nanoscale magnetism, have not been observed yet. 
The diatomic Kagome lattice may have such edge states, but the very existence of them is unclear because there is no chiral symmetry that protects the edge states of graphene \cite{Ryu_2002_3687,Sun_2022_3742,Hu_2022_3748,Hu_2023_3247}. 
Here, we report direct observations of the edge states of a diatomic Kagome lattice, which is made of supramolecular assembly of a triptycene derivative with phenazine moieties (Trip-Phz) on the Pb(111) surface. 
Our scanning tunneling microscopy (STM) observations reveal that Trip-Phz forms well-ordered molecular lattices, terminated by straight boundaries that correspond to the graphene zigzag edge. 
The spectroscopic signatures of the Dirac and flat bands are confirmed using STM and TB calculations. 
Most importantly, we provide direct evidence for the existence of edge states that corresponds to those of graphene.
These states are topological edge states dictated by the quantization of the Zak phase and the bulk-edge correspondence. 
This study provides solid experimental evidence for the edge states in the diatomic Kagome lattice and significantly expands the scope of quantum material research driven by lattice geometry.

\begin{figure}[tb]
\centering
\includegraphics[width=7.5cm]{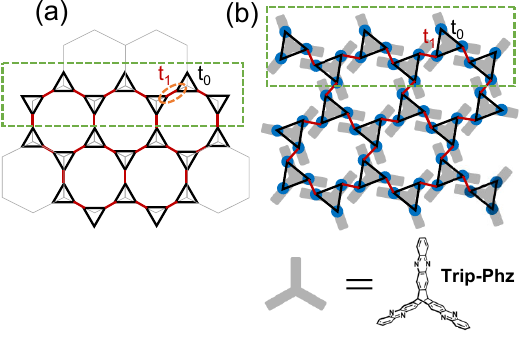}
\caption{Diagram of the diatomic Kagome lattice. (a) Mathematical model with intra-corner coupling $t_0$ and inter-corner coupling $t_1$. The lattice can be regarded as an extension of the honeycomb and Kagome lattices. \textcolor{black}{The dashed orange line indicates a diatomic pair coupled with $t_1$.}(b) Materialization of the diatomic Kagome lattice using Trip-Phz molecules, where $t_0$ and $t_1$ represent intra- and intermolecular couplings, respectively. The gray symbols represent the $\mathrm{C_3}$-symmetric frame of Trip-Phz molecules. Left-handed chirality is chosen for this display. The green rectangles indicate the zigzag-type edge corresponding to that of graphene.
}
\label{Fig1}
\end{figure}

Our strategy for creating the diatomic Kagome lattice in Fig.~1(a) is based on supramolecular chemistry using Trip-Phz \cite{Ushiroguchi_2020_2708,Nemoto_2024_3464}. 
\textcolor{black}{
This type of triptycene derivative molecule comprises three $\pi$-conjugated planes as moieties assembled in a non-planar configuration, featuring a rigid Y-shape with the strict $\mathrm{C_3}$ symmetry. 
}
They can form a highly ordered 2D lattice by $\pi$-$\pi$ pancake bonding between the moieties in a solution and on a substrate surface \cite{Bhola_2013_3712,Kissel_2014_3713,Kohl_2016_3714,Shuku_2018_2707,Ushiroguchi_2020_2708,Das_2021_3376,Grossmann_202_3104,Nemoto_2024_3464}
With the centers of molecules located at the honeycomb nodes and the orbitals of the moieties arranged in a triangular form around them, it can form an ideal diatomic Kagome lattice (Fig.~1(b)). 
This molecular assembly is chiral when formed on a surface; the chirality here is referred to as left-handed, while its mirror-symmetric counterpart as right-handed, in this work \cite{ Nemoto_2024_3464}. 
We note that the chirality mentioned here is geometrical and should not be confused with the chiral symmetry of graphene, which is determined by the structure of the Hamiltonian \cite{Mizoguchi_2019_3152} \textcolor{black}{(see Supporting Note 1)}. 
Figures~1(a) and 1(b) also show basic edge structures of the diatomic Kagome and Trip-Phz lattices, respectively (green rectangles). 
This type of edge is referred to as zigzag edge here because it matches that of the honeycomb lattice in the limit of $t_0 \to \infty$ (i.e., when the triangles are reduced to points). 

\begin{figure*}[tb]
\centering
\includegraphics[width=12cm]{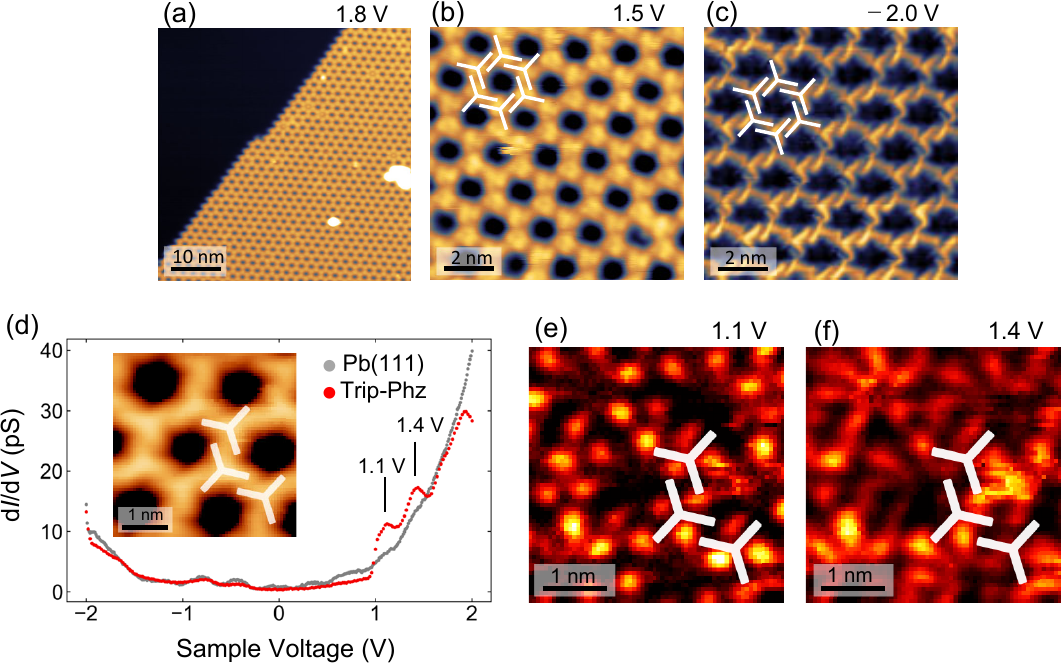}
\caption{STM observations of the diatomic Kagome lattice of Trip-Phz on Pb(111). (a) Large-area STM topographs recorded using $I = 10$ pA and $V = 1.8$ V. (b, c) Detailed topographs acquired using (b) $V = 1.5$ V and (c) $-2.0$ V. (d) $dI/dV$ spectrum averaged over the molecular network of Trip-Phz (red dots) and a reference spectrum on a bare Pb(111) (gray dots). Inset: STM topograph of the relevant area recorded using $V = 2.0$ V, $I = 50$ pA. (e, f) $dI/dV$ images taken over the same area as in (d) using (e) $V = 1.1$ V and (f) $1.4$ V. For the images in (b-f), the molecular adsorption geometries are denoted by white Y-shaped lines.
}
\label{Fig2}
\end{figure*}

We first study the structure of self-assembled molecules using STM (for details, see Supporting Information, Methods). 
Figure~2(a) shows a representative, large-scale, constant-current image of Trip-Phz molecules with an average coverage of $\sim0.3$ monolayers (ML) on a Pb(111) surface. 
We find a monolayer molecular island ($100 \times 80 \ \mathrm{nm^2}$) consisting of a single well-ordered domain. 
Its boundaries are predominantly straight. 
Figures~2(b) and 2(c) present more detailed topographs of the same island, acquired at $V = 1.5 \ \ \mathrm{and}\ -2.0$ V, respectively. 
\textcolor{black}{
These STM images reflect the real-space distribution of the LUMO+1 orbital and the shape of the Trip-Phz molecular framework, respectively; the LUMO+1 states are located at $V = 1.5$ V while there exist no HOMO states at $V= -2.0$ V, as will be revealed below.
} 
The overlaid white Y-shaped lines indicate the molecular frames. 
Evidently, Trip-Phz molecules are arranged in the form of the diatomic Kagome lattice with left-handed chirality, as in Fig.~1(b). 
The lattice parameter of the island was determined to be $1.92\pm 0.04$ nm, and the azimuthal rotation angle of the lattice relative to the principal orientations of the Pb(111) surface $51.3\pm 2 ^\circ$. 
Assuming a commensurate epitaxy on the Pb(111) lattice, the only possible superstructure is Pb(111)-($\sqrt{31}\times \sqrt{31}$)R51$^\circ$, with a lattice constant of $\sqrt{31}a_0 = 1.95$ nm, where $a_0 = 0.350$ nm is the nearest-neighbor distance of the Pb(111) surface. 
We also found a molecular arrangement with the right-handed chirality, which can be denoted as Pb(111)- ($\sqrt{31}\times \sqrt{31}$)R9$^\circ$ (see Supporting Information, Fig.~S1). 

Next, we investigate the spectroscopic features of the ``bulk'' region of an island, sufficiently away from the edges (Fig.~2(d)). 
The STM topograph of this area is shown in the inset, where the locations of the molecules are indicated by Y-shaped lines. 
The $dI/dV$ spectrum averaged over the Trip-Phz lattice (excluding the pore sites) is displayed in red, while a spectrum averaged over a bare Pb(111) surface is plotted in gray as a reference. 
On the positive energy side, two peaks are clearly visible at $V = 1.1 \ \mathrm{and}\ 1.4$ V (indicated by solid lines). 
Since the lowest unoccupied molecular orbital (LUMO) and LUMO+1 of Trip-Phz are located at -2.57 and -2.16 eV below the vacuum level, respectively (Supporting Information, Fig.~S2) \cite{Ushiroguchi_2020_2708}, those peaks can be attributed to molecular bands derived from them. 
\textcolor{black}{
By contrast, on the negative energy side of the spectrum, no HOMO states are discernible down to $V=-2.0$ V.
}
Figures~2(e) and 2(f) display $dI/dV$ maps acquired at $V = 1.1 \ \mathrm{and}\ 1.4$ V, respectively, in the same area. 
For both voltages, the bright spots are located between the three arms of the molecules. 
These spots can be interpreted as the LUMOs of Trip-Phz, since the LUMOs protrude from the $\pi$-conjugated planes of the molecular moieties in the normal direction (Supporting Information, Fig.~S2). 
These states should hardly overlap with those of the Pb(111) surface because they are oriented parallel to the surface \cite{Nemoto_2024_3464}. 
This is advantageous for observing the intrinsic molecular states.
\textcolor{black}{
For discussions on the molecule-substrate interaction, see Supporting Note 2.
}

\begin{figure*}[tb]
\centering
\includegraphics[width=13cm]{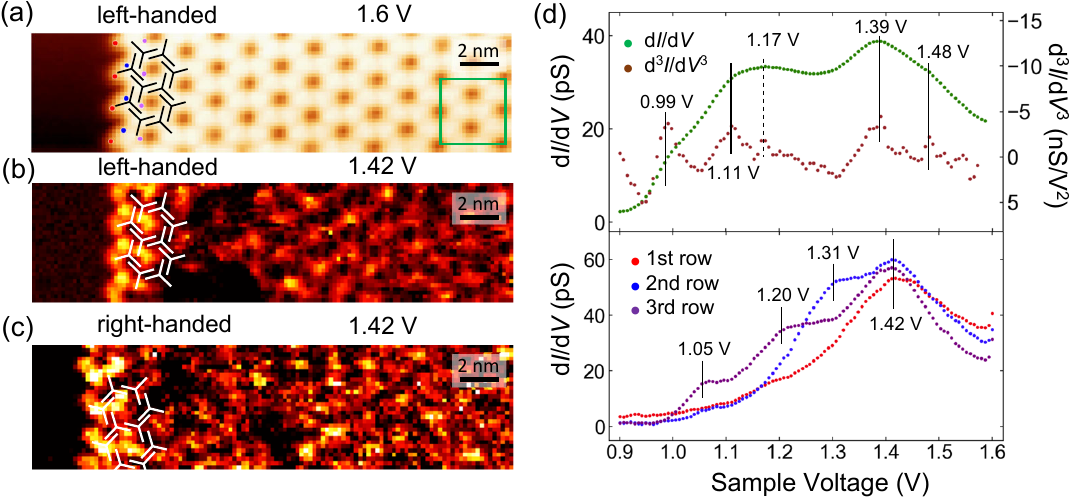}
\caption{Spectroscopic evidence of the zigzag edge states and the Dirac and flat bands of the diatomic Kagome lattice of Trip-Phz. (a) Topograph of the zigzag edge structure with left-handed chirality, recorded using $V = 1.6$ V, $I = 50$ pA. (b) $dI/dV$ image taken at $V = 1.42$ V over the area shown in (a). (c) $dI/dV$ image obtained under the same conditions from a Trip-Phz lattice with right-handed chirality. (d) Upper panel: $dI/dV$ spectrum acquired within the green square in (a) (green dots). The positions of the peak and shoulders were determined from $d^3I/dV^3$ (brown dots), which was obtained by taking the second derivative of $dI/dV$ with respect to V. Lower panel: $dI/dV$ spectra acquired near the edge. The colors of the dots correspond to those in (a); red for the first row (closest to the edge), blue for the second row, and purple for the third row. 
}
\label{Fig3}
\end{figure*}

We now focus on an area including the edge of the Trip-Phz lattice. 
The topograph of Fig.~3(a) reveals that this edge belongs to the zigzag edge defined above, as evident by the overlaid black Y-shaped markers. 
The image also shows that the chirality is left-handed. 
Repeated STM observations of an boundary region confirmed that it was dominated by the zigzag type. 
This is in striking contrast to the honeycomb lattice of graphene, for which armchair edges are predominantly formed \cite{Kobayashi_2005_3678}. 
Figure~3(d) shows $dI/dV$ spectra taken at selected sites within the area of Fig.~3(a). 
The green dots represent the data averaged over an area far from the edge (green square in Fig.~3(a)). 
They reveal that the two peaks identified in Fig.~2(d) actually include four features, as indicated by the solid lines. 
Their positions were determined from $d^3I/dV^3$(brown dots), which was obtained by taking the second derivative of $dI/dV$ with respect to $V$ : $V = 0.99 \ \mathrm{and}\ 1.11$ V for one and $V = 1.39 \ \mathrm{and}\ 1.48$ V for the other. 
An additional weak peak may also exist at $V = 1.17$ V (dotted line). 
These features were repeatedly observed at locations far from the edge (see Supporting Information, Fig.~S3). 
By contrast, the spectra acquired near the edge (red: first row, blue: second row, purple: third row; see also Fig.~3(a)) all exhibit pronounced peaks at $V = 1.42$ V. 
This strongly suggests the presence of edge states centered at $V = 1.42$ V. 
Indeed, a $dI/dV$ image taken at this voltage reveals bright spots localized on the first three rows of the edge, clarifying the spatial distribution of the zigzag edge states (Fig.~3(b)). 
They are observable only around $V = 1.42$ V, in agreement with the pronounced peaks of the $dI/dV$ spectra at this voltage. 
As expected from the symmetry, molecular lattices with right-handed chirality also exhibit edge states, as revealed in the $dI/dV$ image in Fig.~3(c). 
\textcolor{black}{
The weaker appearance of the edge states for this chirality can be attributed to larger spatial fluctuations in $dI/dV$ signal caused by disorder. 
The presence of the edge states at $V = 1.42$ V can also be confirmed in a wider area (see Supporting Information, Fig. S3(c)).
A line profile taken in the normal direction shows the $dI/dV$ intensities at the edge region are significantly larger, on average, than those in the inner bulk area (Fig. S3(d)). 
}
We note that the $dI/dV$ spectra acquired on the second and third rows include shoulders at $V = 1.05, 1.20, \ \mathrm{and}\ 1.31$ V (solid lines). 
Furthermore, $dI/dV$ images acquired around $V = 1.20$ V reveal higher intensities near these rows (see Supporting Information, Fig. S4 and Movie S1). 
They suggest the presence of edge states at these voltages, although not as conspicuous as those observed around $V = 1.42$ V.

To interpret our experimental results, we performed TB calculations. 
The model is based on the diatomic Kagome lattice depicted in Fig.~1, where the intramolecular coupling $t_0$ and intermolecular coupling $t_1$ represent the mutual overlap of the LUMOs of the phenazine moieties. 
Although the substrate is not included in the calculations, it gives only a minor effect on the molecular band structure \cite{Nemoto_2024_3464}. 
The absence of the Shockley surface states on Pb(111) within the relevant energy region also rationalizes this treatment (see Supporting Information, Fig.~S5).
The molecular lattice was terminated by zigzag edges in the $x$ direction, while a periodic boundary condition was imposed in the $y$ direction. 
The transfer integrals $t_0 = -0.141$ eV and $t_1 = -0.096$ eV were determined from the LUMO level splitting of Trip-Phz and a pair of phenazine moieties, respectively, through density functional theory (DFT) calculations (see Supporting Information, Methods).

The left panel of Fig.~4(a) displays the obtained band structures in the $k_y$ direction, which shows the energy dispersions parallel to the edges (blue lines). 
In agreement with previous work, two pairs of Dirac bands ($n=2,3,5,6$) are formed, with the projected Dirac points located at $E = -2.63 \ \mathrm{and}\ -2.17$ eV (indicated by open circles) \cite{Nemoto_2024_3464,Hu_2022_3748,Hu_2023_3247,Shuku_2018_2707}. 
The lower Dirac bands accompany two flat bands ($n=1,4$) at the top and bottom ends. 
These are characteristic of the diatomic Kagome lattice with $t_0/t_1> 2/3$, $t_0, t_1<0$. 
Importantly, because of the termination in the $x$ direction, edge states emerge within some of the band gaps; the states located between the bulk bands $n$ and $n+1$ are denoted as $\mathrm{ES}_{n}$. 
While $\mathrm{ES}_5$ located around $k_y = \pm \pi$ correspond to the edge states of graphene, 
$\mathrm{ES}_1$,$\mathrm{ES}_2$, and $\mathrm{ES}_3$ are distinct because they span different momentum spaces. 
The right panel of Fig.~4(a) shows the corresponding density of states (DOS). 
The edge states appear as sharp peaks (indicated by black arrows) in addition to those due to the van Hove singularity (vHS) within the bulk bands. 
Figure 4(b) shows the spatial distributions of the edge states $\mathrm{ES}_5$ at $k_y=-\pi$ (top panel) and $\mathrm{ES}_2$ at $k_y=0$ (bottom panel) indicated by red dots in Fig.~4(a). 
The diameter of the red circles on the molecular sites is proportional to the local DOS. 
Their spatial dependence shows that $\mathrm{ES}_5$ are localized to the first through third rows of the zigzag edges.
By contrast, $\mathrm{ES}_2$ extend slightly deeper into the interior.  

\begin{figure*}[tb]
\centering
\includegraphics[width=15cm]{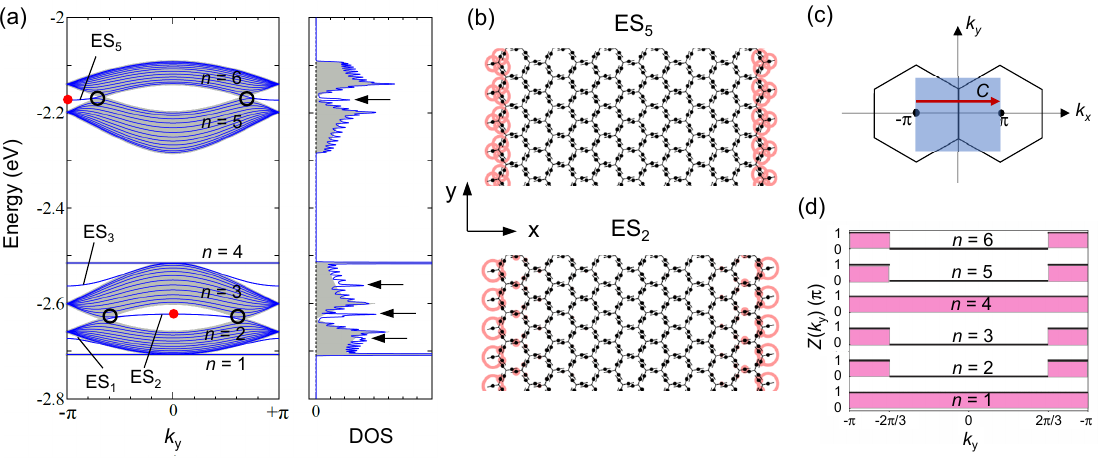}
\caption{Results of tight-binding (TB) calculations for the diatomic Kagome lattice of Trip-Phz. (a) Band structure in the $k_y$ direction (left panel, blue lines) and the corresponding density of states (DOS) (right panel, blue lines). The edge states located between the bulk bands $n$ and $n+1$ are denoted as $\mathrm{ES}_n$. The projected band structure and the corresponding DOS of TB-calculated bulk states are shown by the gray regions. The arrows indicate peaks due to the edge states. (b) Spatial distributions of the edge states of $\mathrm{ES}_5$ at $k_y=-\pi$ (top panel) and  $\mathrm{ES}_2$ at $k_y=0$ (bottom panel) indicated by red dots in (a). Red circles mark the locations of the edge states. The diameter of each circle is proportional to the local DOS of the corresponding states.
(c) Illustration of the Brillouin zones of the diatomic Kagome lattice, where the Zak phase is defined along the path $C$. (d) The Zak phases calculated for the individual bulk bands $n=1,2,\cdots, 6$ as a function of $k_y$. 
}
\label{Fig4}
\end{figure*}

Our spectroscopic data can be largely explained by these theoretical predictions. 
We note that the bands obtained through the TB calculations can be shifted rigidly for comparison with experimental data. 
Regarding the $dI/dV$ spectrum in the central region of the Trip-Phz lattice (Fig.~3(d), green dots), the shoulders at V = 0.99 and 1.11 V can be assigned to the two vHS peaks of the lower Dirac bands ($n=2,3$) and the two flat bands ($n=1,4$). 
Only two peaks out of the four are visible, probably because the proximity of the flat bands to the neighboring vHS peaks and level broadening due to disorder. 
Indeed, the weak feature at $V = 1.17$ V may be assigned to one of the four DOS peaks. 
The peak and shoulder at $V = 1.39 \ \mathrm{and}\ 1.48$ V are attributed to the two vHS peaks in the upper Dirac bands ($n=5,6$). 
Regarding the spectra in the first through third rows (Fig.~3(d), red, blue, and purple dots), the peaks at $V = 1.42$ V are assigned to the edge states $\mathrm{ES}_5$. 
The bright features along the zigzag edges in the $dI/dV$ images (Figs.~3(b) and 3(c)) also correspond well to the spatial distribution of $\mathrm{ES}_5$ (Fig.~4(b), top panel).
By contrast, the assignment of the spectroscopic features observed in the second and third rows at $V = 1.05, 1.20, \ \mathrm{and}\ 1.31$ V is not straightforward. 
We speculate that multiple contributions from the edge states $\mathrm{ES}_1$, $\mathrm{ES}_2$, and $\mathrm{ES}_3$ as well as Dirac and flat bands form the observed spectra.

Here, we show that the edge states found above have a topological origin associated with the Zak phase.
The Zak phase is a Berry phase across an appropriately chosen one-dimensional Brillouin zone \cite{ZakPhase_1D,ZakPhase_graphene}. 
\textcolor{black}{
For an intuitive picture of the Zak phase, refer to Supporting Note 3.
}
In the present case, a path $C$ is chosen at a constant $k_y$ within the 2D Brillouin zone as shown by the red arrow in Fig.~4(c).
The Zak phase $Z_n(k_y)$ for a band $n$ is given by the following equations:
\begin{eqnarray}
Z_n(k_y) &=& \int_{-\pi}^{\pi} A_n(k_x, k_y) dk_x,
\label{eq:ZakPhase1} \\
A_n(k_x, k_y) &=& -i \big\langle u_n(k_x, k_y) |\frac{\partial}{\partial k_x} u_n(k_x, k_y) \big\rangle,
\label{eq:ZakPhase2} 
\end{eqnarray}
where $A_n(k_x, k_y)$ is the Berry connection and $|u_n(k_x, k_y)\rangle$ is the Bloch wave function. 
In the presence of inversion symmetry, $Z_n(k_y)$ is quantized to 0 or $\pi$ (mod $2\pi$), which is valid for the diatomic Kagome lattice.
The presence of the edge states is dictated through the bulk-edge correspondence by the summation of $Z_n(k_y)$ over $n$ 
\begin{equation}
Z(k_y) = \sum_{n=1}^{N} Z_n(k_y),
\label{eq:ZakPhase3}
\end{equation}
where $N$ refers to the number of bands below a relevant energy gap.
If $Z(k_y) =\pi$ (mod $2\pi$), the system is topologically nontrivial and the edge states appear in this gap. 
Otherwise, \textit{i.e.} if $Z(k_y) =0$ (mod $2\pi$), the edge states are absent. 
Figure~4(d) shows the Zak phase calculated for the diatomic Kagome lattice with $t_0 = -0.141$ eV and $t_1 = -0.096$ eV, demonstrating the quantization to 0 or $\pi$.
The summation over $n=1, 2, \cdots, 5$ leads to $Z(k_y) =\pi$ (mod $2\pi$) for $2\pi/3<|k_y|<\pi$ and 0 for elsewhere.
This exactly corresponds to the presence of the edge states $\mathrm{ES}_5$ within $2\pi/3<|k_y|<\pi$. 
The other edge states can also be dictated in the same manner.
Thus, the edge states found here have a topological origin associated with the quantization of the Zak phase, which dictates their existence through the bulk-edge correspondence.
They can be regarded as a generalization of the zigzag edge states of graphene in a system without the chiral symmetry.

Recently, Hu et al. predicted through TB calculations that edge states exist within an energy gap for certain types of the diatomic Kagome lattice \cite{Hu_2023_3247,Hu_2022_3748}. 
They also showed the existence of corner states as second-order topological states in the same energy gap. 
Both of them were observed by $dI/dV$ imaging using STM. 
Our TB parameters $t_0/t_1> 2/3$, $t_0, t_1<0$ preclude such states, but nevertheless, it will be worth searching for them by tuning molecular parameters. 
Indeed, the ratio $t_0/t_1$ is sensitive to the molecular lattice constant, which in turn is tunable by choosing an appropriate substrate \cite{Nemoto_2024_3464}.
Telychko et al. and Yin et al. reported the formation of diatomic Kagome lattices by confining the Shockley surface states of noble metals using molecular templates \cite{Telychko_2021_3094,Yin_2024_3409}. 
The latter authors also observed edge states around the boundary of a molecular island in the gap region of a breathing Kagome lattice. 
The edge states observed in the present work are derived from molecular orbitals forming an intrinsic tight-binding lattice. 
They are connected to the projected Dirac points, and therefore have a different origin.

Finally, we remark on the prospects of the present study. 
Recently, Kagome-related materials have been one of the hottest topics in condensed-matter physics \cite{Haldane_1988_3007,Kane_2005_1682,Novoselov_2005_184,Tang_2011_2934,Ye_2018_2941,Li_2018_3733,Ortiz_2020_3751,Jiang_2021_2938}
For this purpose, various layered inorganic materials have been synthesized, but it still remains challenging to design and find appropriate systems. 
This is because the complicated interactions between multiple atomic sites and orbitals often destroy the electronic properties predicted by theoretical models \cite{Jovanovic_2022_3753}. 
By contrast, as demonstrated here, it is possible to fabricate a nearly ideal 2D lattice using chemically designed non-planar $\pi$-conjugated molecules as building blocks. 
The upright configuration of the moiety planes enhances intralayer coupling while suppressing coupling with the substrate and neighboring stacking layers. 
The use of other non-planar $\pi$-conjugated molecules also appears promising; for example, triptycene trithianthrene (TTA) forms a layered crystal with a breathing Kagome lattice in a solution \cite{Shuku_2023_3286,Tang_2022_3662}. 
This will be the subject of a forthcoming study.

In summary, we successfully fabricated a highly ordered diatomic Kagome lattice through the self-assembly of Trip-Phz on Pb(111), which was predominantly terminated by zigzag-type edges. 
Our STM measurements and TB calculations provided direct evidence for the zigzag edge states that correspond to those of graphene within a pair of Dirac bands. 
They are topological states given by the quantization of the Zak phase and the bulk-edge correspondence.
This work offers a new platform based on supramolecular technology for exploring quantum materials with unique lattice geometries.

\nocite{Nemoto_2024_3464,Gaussian16_3355,Kresse_VASP1,Kresse_VASP2,Bloechl_1994_3336,Perdew_1996_3337,Hamada_2014_3338,Monkhorst_1976_3339}

\vspace{10pt}

\section*{Associated Content}
\subsection*{Supporting information}
The Supporting Information is available online. Contents: Preparation and purification of Trip-Phz, sample preparation and STM measurements, tight-binding and DFT calculations, and references (Methods); the bias voltage dependence of high-resolution STM topographs of a Trip-Phz island (Fig.~S1), the energy levels and spatial distributions of the LUMOs (Fig.~S2), $dI/dV$ spectra obtained in a bulk region of a Trip-Phz lattice (Fig.~S3), the edge states of Trip-Phz lattices observed in the same area as in Figs.~3(a) and 3(b)(Fig.~S4), and Electronic band structure of Pb(111) (Fig.~S5); Caption for Supporting Movie.

\section*{Author Information}
\subsection*{Corresponding Authors}
\begin{itemize}
\item Alexander Weismann -- Institut f\"ur Experimentelle und Angewandte Physik, Christian-Albrechts-Universit\"at, 24098 Kiel, Germany; ORCID iD: 0000-0003-2487-3917 \\
Email: weismann@physik.uni-kiel.de
\item  Rie Suizu -- Synchrotron Light Application Center, Saga University, Honjo, Saga, 840-8502, Japan; Department of Chemistry and IRCCS, Nagoya University, Furo-cho, Chikusa-ku, Nagoya 464-8602, Japan; Japan Science and Technology Agency (JST), PRESTO, 4-1-8 Honcho, Kawaguchi, Saitama, 332-0012, Japan; ORCID iD: 0000-0001-7632-2186  \\
Email: rsuizu@cc.saga-u.ac.jp
\item Takashi Uchihashi -- Research Center for Materials Nanoarchitectonics (MANA), National Institute for Materials Science, 1-1, Namiki, Tsukuba, Ibaraki 305-0044, Japan; Graduate School of Science, Hokkaido University, Kita-10 Nishi-8, Kita-ku, Sapporo 060-0810, Japan; ORCID iD: 0000-0003-0811-5665 \\
Email: UCHIHASHI.Takashi@nims.go.jp
 \end{itemize}

\subsection*{Authors}
\begin{itemize}
\item Ryohei Nemoto -- Research Center for Materials Nanoarchitectonics (MANA), National Institute for Materials Science, 1-1, Namiki, Tsukuba, Ibaraki 305-0044, Japan; Department of Physics, School of Science, Institute of Science Tokyo, 2-12-1, Ookayama, Meguro-ku, Tokyo 152-8551, Japan; ORCID iD: 0000-0002-7343-7268
\item Xiangzhi Meng -- Institut f\"ur Experimentelle und Angewandte Physik, Christian-Albrechts-Universit\"at, 24098 Kiel, Germany; ORCID iD: 0000-0001-7887-4240
\item Behzad Mortezapour -- Institut f\"ur Experimentelle und Angewandte Physik, Christian-Albrechts-Universit\"at, 24098 Kiel, Germany
\item Saya Nakano -- Department of Physics, Nara Women's University, Kitauoyanishi-machi, Nara 630-8506, Japan
\item Masahisa Tsuchiizu -- Department of Physics, Nara Women's University, Kitauoyanishi-machi, Nara 630-8506, Japan; ORCID iD: 0009-0002-2975-8096
\item Ryuichi Arafune -- Research Center for Materials Nanoarchitectonics (MANA), National Institute for Materials Science, 1-1, Namiki, Tsukuba, Ibaraki 305-0044, Japan; ORCID iD: 0000-0003-4371-6116
\item Noriaki Takagi -- Graduate School of Human and Environmental Studies, Kyoto University, Kyoto 606-8501, Japan; ORCID iD: 0000-0002-0799-9772
\item Sho Nakamura -- School of Science and Technology, Kwansei Gakuin University, Sanda, Hyogo 669-1330, Japan
\item Katsunori Wakabayashi -- Research Center for Materials Nanoarchitectonics (MANA), National Institute for Materials Science, 1-1, Namiki, Tsukuba, Ibaraki 305-0044, Japan; School of Science and Technology, Kwansei Gakuin University, Sanda, Hyogo 669-1330, Japan; ORCID iD: 0000-0002-9147-9939
\item Richard Berndt -- Institut f\"ur Experimentelle und Angewandte Physik, Christian-Albrechts-Universit\"at, 24098 Kiel, Germany; ORCID iD: 0000-0003-1165-9065
\item Kunio Awaga -- Department of Chemistry and IRCCS, Nagoya University, Furo-cho, Chikusa-ku, Nagoya 464-8602, Japan; National Institute of Technology (KOSEN), Toyota College, 2-1 Eiseicho, Toyota, Aichi 471-8525, Japan; ORCID iD: 0000-0002-2193-0747
 \end{itemize}

\subsection*{Note}
The authors declare no competing financial interest.

\section*{Acknowledgments}
This work was supported financially by JSPS KAKENHI (Grant Numbers 25H00867, 25K01670, 20H05621, 20H02707, 23K03322, 25K01609, 22H05473), Japan Science and Technology Agency (JST) PRESTO Grant JPMJPR21A9, and World Premier International Research Center (WPI) Initiative on Materials Nanoarchitectonics, MEXT, Japan.


\bibliography{MyEndNoteLibrary2}

\newpage

\section*{TOC Graphic}
\begin{figure}[h!]
\centering
\includegraphics[width=8.25 cm]{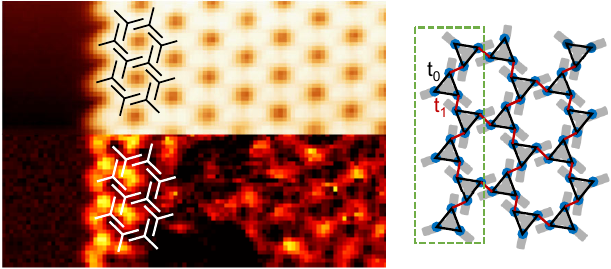}
\end{figure}

\end{document}